\documentclass[12pt]{article}
\usepackage{amsmath}
\usepackage[psamsfonts]{amssymb}

\title{All Teleportation and Dense Coding Schemes}
 \author{R.~F. Werner\thanks{Electronic Mail: \texttt{r.werner@tu-bs.de}}
  \\[1ex]
  {\small Institut f{\"u}r Mathematische Physik, TU Braunschweig,}\\
  {\small Mendelssohnstr.3, 38106 Braunschweig, Germany.}}
\date{March 16, 2000}

\def\H{{\cal H}}
\def\B{{\cal B}}
\def\K{{\cal K}}

\def\Cx{{\mathbb C}} 
\def\idty{{\rm 1\mkern -4.8mu I}}
\def\tr{\mathop{\rm tr}\nolimits}
\def\id{{\rm id}}

\def\ket #1{\vert #1\rangle}
\def\braket #1#2{\langle #1 \vert #2\rangle}
\def\ketbra #1#2{\vert #1\rangle\!\langle #2\vert}
\def\abs #1{\vert#1\vert}
\def\norm #1{\Vert#1\Vert}
\def\det #1{{\mathop{\rm det} \! \left(#1\right)}}
\def\proof{\par\noindent{\it Proof:\ }}
\def\QED{\leavevmode\unskip\penalty9999 \hbox{}\nobreak\hfill
     \quad\hbox{\leavevmode  \hbox to.77778em{%
               \hfil\vrule   \vbox to.675em%
               {\hrule width.6em\vfil\hrule}\vrule\hfil}}
     \par\vskip24pt}

\newtheorem{The}{Theorem} \newtheorem{Def}[The]{Definiton}
\newtheorem{Lem}[The]{Lemma} \newtheorem{Pro}[The]{Proposition}
\newtheorem{Cor}[The]{Corollary}

\begin{document}
\maketitle

\abstract{We establish a one-to-one correspondence between (1)
quantum teleportation schemes, (2) dense coding schemes, (3)
orthonormal bases of maximally entangled vectors, (4) orthonormal
bases of unitary operators with respect to the Hilbert-Schmidt
scalar product, and (5) depolarizing operations, whose Kraus
operators can be chosen to be unitary. The teleportation and dense
coding schemes are assumed to be ``tight'' in the sense that all
Hilbert spaces involved have the same finite dimension $d$, and
the classical channel involved distinguishes $d^2$ signals. A
general construction procedure for orthonormal bases of unitaries,
involving Latin Squares and complex Hadamard Matrices is also
presented. }

\section{Introduction}

Teleportation and dense coding are two processes, which stood at
the beginning of modern Quantum Information Theory. They both
demonstrated radically new features of quantum information as
opposed to classical information, in that both would be impossible
without the assistance of entangled states. Indeed, the attempt of
using the properties of a classically correlated system shared by
sender and receiver to improve the transmission rate of a
classical channel can easily be seen to be hopeless. But this is
precisely what happens in teleportation and dense coding, and
dramatically so, because without entanglement assistance
teleportation, i.e., the transmission of quantum information on a
classical channel, would not only be less efficient, but virtually
impossible.

In the original papers \cite{densecod,tele} the new possibilities were
demonstrated by giving an explicit example, based on qubits. It
was clear early on that extensions to systems with higher
dimensional Hilbert spaces were possible, not only to powers of
$2$, by running the process several times, but to any dimension
$2\leq d<\infty$ \cite{tele}.

The task set in this paper is to do this systematically, and to
classify {\it all} schemes for teleportation and dense coding.
There are several reasons for doing this. The first is, of course,
to take these miracle machines apart and to analyze what makes
them work: what is the mathematical structure one really needs to
set up such a scheme? For the present author one motivation of
this kind was to understand the surprising observation that each
of the published teleportation schemes also works as a dense
coding scheme, and conversely: sender Alice and receiver Bob
merely have to swap the equipment they use. An attempt at a direct
proof of this failed, and indeed, as discussed below, the
statement fails in general, but is true in the special case of
``tight'' schemes.

The second reason for attempting a complete classification of
teleportation schemes is more practical. In spite of amazing
progress in recent years, experiments in quantum information
processing are still quite difficult. Hence, for realizing a
teleportation scheme it is useful to have a systematic overview of
the options, before going on to find the one which is the easiest
to implement. This also goes for approximate realizations. And in
order to find feasible approximate teleportation schemes it is
probably once again necessary to understand the manifold of exact
realizations.

The aim of determining all schemes is not quite achieved in this
paper, in two respects. Firstly, we will only look at the case
when dense coding and teleportation are realized optimally with
minimal resources, in the sense of Hilbert space dimensions and
number of distinguishable classical signals. As in the well-known
qubit case, this means that an entangled state between systems of
the same dimension $d$ as the input systems is used, and the
classical channel distinguishes $d^2$ signals. That is, the
classical capacity of the quantum channel is exactly doubled by
dense coding, and teleportation requires twice as much classical
channel capacity as the quantum capacity of the channel set up by
this scheme. We will call schemes with these dimension parameters
{\it tight}. As mentioned above, for these dimensions the symmetry
between teleportation and dense coding holds perfectly.
Classifying all schemes beyond the tight case appears to be more
difficult because there is too much freedom, which cannot be
parametrized in a simple way (see, however, \cite{braun}).

The second respect in which this paper falls short of a complete
classification is that we can only reduce it to another
``standard'' problem, namely the construction of orthonormal bases
of unitary operators with respect to the scalar product
$(A,B)\mapsto d^{-1}\tr(A^*B)$. In the last section we provide a
fairly general construction for such bases. However, even this
construction has to rely on other well-known but not completely
classified combinatorial designs, namely Latin squares, and
complex Hadamard matrices. This suggests that a complete
construction procedure for all unitary bases would be at least as
difficult as a complete classification of Latin squares or
Hadamard matrices, and hence hardly a promising task.

The paper is organized as follows:  in Section~2 the Main Theorem
is stated: an equivalence in the tight case between teleportation
schemes, dense coding schemes, orthonormal unitary bases, bases of
maximally entangled vectors, and so-called unitary depolarizers.
Basic consequences of the Theorem are discussed. Section~3
contains the proof, divided into subsections, each devoted to some
implication in the big equivalence. In writing the proof an
attempt was made to include also simple steps explicitly, and to
make as transparent as possible why the tightness condition is
crucial. Finally, in Section~4 we present the ``Shift and
Multiply'' construction of unitary bases, which are then
classified in terms of Latin squares and Hadamard matrices.

\section{Main result}

In order to state our result we use the following notation and
terminology: When $\H$ is a Hilbert space, we denote by $\B(\H)$
the space of bounded linear operators on $\H$. A {\it channel}
converting quantum systems with Hilbert space $\H_{\rm in}$ into
systems with Hilbert space $\H_{\rm out}$ is a linear operator
$T:\B(\H_{\rm out})\to\B(\H_{\rm in})$, which is completely
positive \cite{davies,paulsen} and normalized as $T(\idty)=\idty$.
A (discrete)
{\it observable} $F$ on $\H$ over an output parameter space $X$ is
a collection of positive operators $F_x\in\B(\H)$ such that
$\sum_xF_x=\idty$. A {\it density operator} on $\H$ is a positive
operator with trace $1$. The basic {\it probabilistic
interpretation} of these objects is fixed by the prescription that
$\tr(\omega T(F_x))$ is the probability to get the measuring
result ``$x$'' on systems prepared according $\omega$, before
passing through the channel $T$. Finally, we call a vector
$\Psi\in\H\otimes\H$ {\it maximally entangled}, if it is
normalized, and its reduced density operator is maximally mixed,
i.e., a multiple of $\idty$:
\begin{equation}\label{maxent}
  \braket\Psi{(A\otimes\idty)\Psi}=(\dim\H)^{-1}\tr(A)\;.
\end{equation}

Let us set up the equations describing dense coding and
teleportation in this language. In both cases, the beginning of
each transmission is to distribute the parts of an entangled state
$\omega$ between sender Alice and receiver Bob. Only then Alice is
given the message she is supposed to send, which is a quantum
state in the case of teleportation and a classical value in case
of dense coding. She codes this in a suitable way, and Bob
reconstructs the original message by evaluating Alice's signal
jointly with his entangled subsystem. For {\em dense coding},
assume that $x\in X$ is the message given to Alice. She encodes it
by transforming her entangled system by a channel $T_x$, and
sending the resulting quantum system to Bob, who measures an
observable $F$ jointly on Alice's particle and his. The
probability for getting $y$ as a result is then $\tr\bigl(\omega
(T_x\otimes\id)(F_y)\bigr)$, where the ``$\otimes\id$'' expresses
the fact that no transformation is done to Bob's particle while
Alice applies $T_x$ to hers. If everything works correctly, this
expression has to be $1$ for $x=y$, and $0$ otherwise (see eq.
(\ref{densecod})).

Let us take a similar look at {\it teleportation}. Here three
quantum systems are involved: the entangled pair in state
$\omega$, and the input system given to Alice, in state $\rho$.
Thus the overall initial state is $\rho\otimes\omega$. Alice
measures an observable $F$ on the first two factors, obtaining a
result $x$ sent to Bob. Bob applies a transformation $T_x$ to his
particle, and makes a final measurement of an observable $A$ of
his choice. Thus the probability for Alice measuring $x$ and for
Bob getting a result ``yes'' on $A$, is
$\tr(\rho\otimes\omega)(F_x\otimes T_x(A))$. Note that the tensor
symbols in this equation refer to different splittings of the
system ($1\otimes23$ and $12\otimes3$, respectively).
Teleportation is successful, if the overall probability for
getting $A$, computed by summing over all possibilities $x$, is
the same as for an ideal channel, i.e., $\tr(\rho A)$, as in eq.
(\ref{telepo}).

The only relationship between the Hilbert spaces involved, which
this description requires, is that the input and output spaces of
the teleportation line are the same, since the whole teleportation
process is equivalent to the identity. In some sense the best
results (minimal dimension for the Hilbert spaces carrying the
entangled state, best ratio of achieved capacity to capacity used)
are obtained in the special case, where all Hilbert spaces have
the same dimension $d$, and exactly $\abs X=d^2$ signals are
distinguished. We call this the {\em tight} case, and the main
Theorem refers only to this case.

\begin{The} Let $\H$ be a $d$-dimensional Hilbert space
($d<\infty$), and $X$ a set of $d^2$ elements. Consider the
following types of objects:
\begin{enumerate}
\item {\bf Teleportation schemes}
consisting of \begin{itemize}
   \item a density operator $\omega$ on $\H\otimes\H$
   \item a collection of channels $T_x:\B(\H)\to\B(\H)$, $x\in X$
   \item an observable $F_x$, $x\in X$ on $\H\otimes\H$
   \end{itemize}
such that, for all density operators $\rho$ on $\H$, and
$A\in\B(\H)$:
\begin{equation}\label{telepo}
  \sum_{x\in X}\tr(\rho\otimes\omega)(F_x\otimes T_x(A))
   =\tr\rho A \;.
\end{equation}
\item{\bf Dense coding schemes}, consisting of the same objects
as a Teleportation Scheme, but satisfying, instead of
(\ref{telepo}), the equation
\begin{equation}\label{densecod}
  \tr\bigl(\omega (T_x\otimes\id)(F_y)\bigr)=\delta_{xy}\;.
\end{equation}
\item{\bf Bases of maximally entangled vectors}, i.e. families of
maximally entangled vectors $\Phi_x\in\H\otimes\H$, $x\in X$ such
that
\begin{equation}\label{entbasis}
 \braket{\Phi_x}{\Phi_y}=\delta_{xy}\;
\end{equation}
\item{\bf Bases of unitary operators}, i.e., collections of
unitary operators $U_x\in\B(\H)$, $x\in X$ such that
\begin{equation}\label{unibasis}
  \tr(U_x^*U_y)=d\,\delta_{xy}\;.
\end{equation}
\item{\bf Unitary depolarizers},i.e., collections of
unitary operators\hfill\break%
$U_x\in\B(\H)$, $x\in X$ such that for any
$A\in\B(\H)$:
\begin{equation}\label{unidep}
 \sum_x U_x^*AU_x=d\,\tr(A)\,\idty\;.
\end{equation}
\end{enumerate}
Then, given any object of any one of these types, one can
construct an object of each of the types, using the following
equations:
\begin{eqnarray}
\label{O2omega}   \omega&=&\ketbra\Omega\Omega,\quad\text
                          {with $\Omega$ maximally entangled.}\\
\label{Phi2F}        F_x&=&\ketbra{\Phi_x}{\Phi_x}\\
\label{U2T}       T_x(A)&=&U_x^*AU_x\\
\label{U2Phi}     \Phi_x&=&(U_x\otimes\idty)\Omega\;.
\end{eqnarray}
\end{The}

The logical structure of this result is maybe slightly unusual, so
we begin by giving some examples how it is used. We can use it,
for example, as a construction procedure: once we are given a
unitary basis, we can get from the equations (\ref{O2omega}) to
(\ref{U2Phi}) a teleportation scheme and a dense coding scheme.
Moreover, since we could also start with these schemes, ending up
with the unitary basis we are assured that {\it every}
teleportation or dense coding scheme is obtained in this way,
i.e., this construction is exhaustive. In particular, we learn
that any tight teleportation scheme is necessarily of a very
special form: the entangled state $\omega$ must be pure and
maximally entangled, the channels $T_x$ must be unitarily
implemented, and the observable $F$ must be a complete von Neumann
measurement.

Another result contained in this Theorem is the amazing
equivalence between (1) and (2): any teleportation scheme works as
a dense coding scheme and conversely. Alice and Bob merely have to
swap their equipment to convert one into the other. We must
emphasize, however, that the tightness condition is absolutely
crucial for this equivalence. For simplicity, we will discuss this
only in the case that $\abs X=n$ is not fixed to be $d^2$, leaving
aside the more difficult question what kind of trade-off between
resources becomes possible, when $\omega$ lives on
$\H_1\otimes\H_2$, with dimensions other than $d\otimes d$.

The basic difference between teleportation and dense coding is
that the parameters $d$ and $n$ have opposite roles: For
teleportation $d$ describes the size of the signal to be sent, and
$n$ describes a resource, so the problem becomes more difficult
when we increase $d$ and decrease $n$. For dense coding, it is
exactly the opposite. Therefore, it is easy to show that
teleportation (resp.\ dense coding) schemes exist whenever $n\geq
d^2$ (resp.\ $n\leq d^2$). In fact, for teleportation one can take
$X$ to be a continuum, and replace the sum in the teleportation
equation by an integral \cite{braun}, but the dense coding
equation would make no sense then. The optimality of these
dimension inequalities, i.e., that no teleportation (resp.\ dense
coding) scheme exists with $n<d^2$ (resp.\ $n> d^2$), is also a
corollary of Theorem~1. To prove it, suppose we had a
teleportation scheme with $n<d^2$. Then we could add $(d^2-n)$
irrelevant classical signals happening with probability zero
($F_x=0$), and apply the Theorem, which says that all $F_x$ must
be non-zero after all. The same reasoning works for dense coding
with the operation of throwing in a few unused Hilbert space
dimensions.

Of course, our Theorem is efficient as a construction procedure
for dense coding and teleportation schemes only to the extent that
unitary bases can be generated. After giving the proof of the
Theorem, we will therefore describe the most general construction
for such bases known to us.

\section{Proof of Theorem 1}

\subsection*{Proof of the Implications ``3$\Longleftrightarrow$4''}
Implicit in the formulation of the Theorem is the claim that the
equation (\ref{U2Phi}) $\Phi_x=(U_x\otimes\idty)\Omega$ not only
determines $\Phi_x$ in terms of $U_x$ but also, conversely,
determines $U_x$ in terms of $\Phi_x$. This connection is based on a
general construction, by which the $d^2$ matrix elements of an
operator $A:\H\to\H$ are identified with the $d^2$ components of a
vector $\Psi$. This identification depends on the choice of a
maximally entangled vector $\Omega$. By choosing appropriate
orthonormal bases $e_k$, $k=1,\ldots,d$ in the first and second
tensor factor, such a vector can be written in ``Schmidt form''
as
\begin{equation}\label{Schmidt}
  \Omega=\frac1{\sqrt d}\sum_k e_k\otimes e_k\;.
\end{equation}
Then a one-to-one correspondence between operators $A\in\B(\H)$
and $\Psi\in\H\otimes\H$ is given by the equation
$\braket{e_k}{Ae_\ell}=\sqrt d\;\braket{e_k\otimes e_\ell}\Psi$.
We will use this in the form
\begin{equation}\label{A2Psi}
  \Psi=(A\otimes\idty)\Omega=(\idty\otimes A^T)\Omega\;,
\end{equation}
where the transpose operation $A\mapsto A^T$ is defined in the
basis $e_k$. Then if $A$ and $\Psi$ and, similarly, $A'$ and
$\Psi'$ are related in this way,
\begin{equation}\label{kkkk}
  \braket{\Psi}{(B\otimes\idty)\Psi'}=\frac1d\tr(A^*BA')\;,
\end{equation}
for arbitrary $B\in\B(\H)$. Thus $\Psi$ is maximally entangled iff
this expression (for $A=A'$) is equal to $d^{-1}\tr(B)$, i.e., iff
$A$ is unitary. Moreover, setting $B=\idty$, the scalar product of
vectors $\Psi,\Psi'$ is translated to $d^{-1}\tr(A^*A')$ in terms
of $A,A'$. Taking all this together, we get the one-to one
correspondence between unitary bases and bases of maximally
entangled vectors, as claimed. Note, however, that this
correspondence depends on the choice of the reference maximally
entangled vector $\Omega$.

\subsection*{Proof of the Implications ``4$\Longleftrightarrow$5''}
This proof is relatively straightforward, since we are talking
about only one type of objects, collections of $d^2$ unitaries
$U_x\in\B(\H)$. It is, however, also a crucial step for the entire
proof, since it is here that the consequences of the tightness
condition are seen. We will prove this in a form, which is also
needed later to establish that the state $\omega$ in teleportation
and dense coding schemes is necessarily maximally entangled.

The basic observation concerning matching dimensions is the
following.

\begin{Lem}\label{l.complete} $D$ vectors $\phi_1,\ldots,\phi_D$ in a
$D$-dimensional Hilbert space form an orthonormal basis if and
only if
\begin{equation}\label{e.complete}
    \sum_{k=1}^D\ketbra{\phi_k}{\phi_k}=\idty\;.
\end{equation}
\end{Lem}

Of course, this is false when there are more vectors than the
dimension of the Hilbert space. Such families of vectors are
called ``overcomplete''. They exist and are an interesting mathematical
structure of their own. On the other hand, fewer vectors than the
dimension can never satisfy eq. (\ref{e.complete}), because the
rank (dimension of the range) of the operator on the left hand
side is at most the number of vectors.

\vskip20pt

\proof It is a well-known fact that eq. (\ref{e.complete}) holds
for any orthonormal basis. Conversely, we find from eq.
(\ref{e.complete}) that, for each $k$,
$\ketbra{\phi_k}{\phi_k}\leq\idty$, which is the same as
$\norm{\phi_k}^2\leq1$. On the other hand, taking the trace of
(\ref{e.complete}), we get $\sum_k\norm{\phi_k}^2=\tr(\idty)=D$.
This is only possible, when $\norm{\phi_k}^2=1$ for all $k$. Hence
the operators $\ketbra{\phi_k}{\phi_k}$ are hermitian projections,
and we can invoke the observation that hermitian projections
$p_1,p_2$ with $p_1+p_2\leq\idty$ are necessarily orthogonal. (For
a quick proof, sandwich the inequality between factors $p_1$,
finding $p_1+p_1p_2p_1\leq p_1$, i.e.,
$p_1p_2p_1=(p_2p_1)^*(p_2p_1)\leq0$, and hence $p_2p_1=0$). \QED

We now apply this Lemma to a collection of $D=d^2$ operators in
$\B(\H)$, where this space is considered as a Hilbert space with a
suitable scalar product.

\begin{Pro}Consider $d^2$ operators $K_1,\ldots,K_{d^2}$ on a
$d$-dimen\-sional Hilbert space $\H$, and let $R >0$ be an
invertible operator on $\H$.\\ Then the following conditions are
equivalent
\begin{enumerate}
\item $\tr(K_x^*R ^{-1}K_y)=\delta_{xy}$, for $x,y=1,\ldots,d^2$
\item $\sum_xK_x^*CK_x=\tr(R  C)\idty$ for all $C\in\B(\H)$.
\end{enumerate}
\end{Pro}

\proof Let us define a scalar product $\braket\cdot\cdot_R $ on
$\B(\H)$ by
\begin{equation}\label{sp-rho}
  \braket AB_R =\tr(A^*R ^{-1}B)\;.
\end{equation}
Since $R $ is positive and invertible, this is indeed a scalar
product, satisfying $\braket AA_R =0$ only for $A=0$. Condition~1
then simply says that the $K_x$ are an orthonormal basis. By the
previous Lemma this is equivalent to the completeness relation
(\ref{e.complete}), so all we have to do is to show that this
relation, adapted to the special scalar product at hand, is
equivalent to Condition~2 of the present Lemma. The completeness
relation is that, for any $A,B\in\B(\H)$,
\begin{equation}\tag{$\ast$}
  \braket AB_R
     =\sum_x\braket A{K_x}_R \;\braket{K_x}B_R \;.
\end{equation}
It suffices to evaluate this on rank one operators $A,B\in\B(\H)$,
since these span the whole space. We take
$A=\ketbra{\phi_1}{\phi_2}$ and $B=\ketbra{\psi_1}{\psi_2}$. Then
the left hand side of equation~($\ast$) becomes
\begin{equation}\tag{$\ast$LHS}
 \braket{\phi_1}{R ^{-1}\psi_1}\braket{\psi_2}{\phi_2}\;,
\end{equation}
whereas the right hand side is
\begin{equation}\tag{$\ast$RHS}
  \sum_x \braket{\phi_1}{R ^{-1}K_x \phi_2}
         \braket{\psi_2}{K_x^*R ^{-1} \psi_1}
  =\braket{\psi_2}{M\phi_2}\;,
\end{equation}
with
\begin{displaymath}
    M=\sum_x K_x^*R ^{-1}\ketbra{\psi_1}{\phi_1}R ^{-1}K_x
     \equiv\sum_x K_x^*CK_x\;
\end{displaymath} where we have interchanged the two factors in
each term, and introduced the abbreviation $C$. Since
($\ast$LHS)=($\ast$RHS) for every ${\psi_2},{\phi_2}$, we find
$M=\braket{\phi_1}{R^{-1}\psi_1}\;\idty$. The factor is readily
identified as
\goodbreak
$\braket{\phi_1}{R ^{-1}\psi_1}=\tr(R
C)$. Since operators of the form $C$ span $\B(\H)$, the
completeness relation thus becomes equivalent to
$\sum_xK_x^*CK_x=\tr(R C)\idty$ for all $C$, which completes the
proof. \QED

The special case of this Proposition, where each $K_x$ is unitary
and $R =\frac1d\idty$, is exactly the relationship between items 4
and 5 of Theorem~1. However, there is another consequence needed
later on:

\begin{Cor}\label{noUrhobasis}
Let $U_1,\ldots,U_{d^2}\in\B(\H)$ be unitaries in a
$d$-dimensional Hilbert space $\H$, and $\rho$ a density operator
such that $\tr(U_x^*\rho U_y)=\delta_{xy}$. Then
$\rho=d^{-1}\idty$.
\end{Cor}

\proof Since the $U_x$ are an orthonormal set whose cardinality is
the dimension, there can be no null vectors of this scalar
product, i.e., $\tr(A^*\rho A)=0$ implies $A=0$. Hence $\rho$ is
invertible, and we can apply the previous Proposition  with
$R=\rho^{-1}$,  finding that
$\sum_xU_x^*AU_x=\tr(\rho^{-1}A)\idty$. The trace of this equation
is  $d^2\tr(A)=d\tr(\rho^{-1}A)$. This holds for all $A$, i.e.,
$\rho^{-1}=d\idty$. \QED

\subsection*{Proof of 
``(3 or 4)$\Longrightarrow$(1 and 2)''}

Suppose now we are given either a basis of unitary operators or of
maximally entangled vectors. Then we can choose a maximally
entangled vector $\Omega$ and use equation~(\ref{U2Phi}) as in in
the proof of  ``3$\Leftrightarrow$4'' to define the other kind of
basis. Equations (\ref{Phi2F}) and (\ref{U2T}) then become
explicit definitions of the observable $F_x$ and the
transformations $T_x$, respectively, so all the objects needed for a
teleportation or dense coding scheme are defined, and we only need
to verify that equations~(\ref{telepo}) and (\ref{densecod}) are
indeed satisfied.

In the teleportation equation an expectation value is generated
between a state on the first and an observable on the third factor
of a triple tensor product. This is a consequence of a similar
``teleportation equation'' on the level of vectors, which we now
state. For later use we prove a certain converse at the same time.

\begin{Lem}\label{Lem:scprod}
Let $\Omega\in\Cx^d\otimes\Cx^d$ be the maximally
entangled vector $\Omega=d^{-1/2}\sum_ke_k\otimes e_k$, where
$e_k$, $k=1,\ldots,d$ is the standard basis of $\Cx^d$. Let
$M\in\B(\Cx^d)$, and $\mu\in\Cx$. Then the equation
\begin{displaymath}
    \braket{\phi\otimes\Omega}
          {(\idty\otimes
          M\otimes\idty)\Omega\otimes\psi}=\mu\braket\phi\psi
\end{displaymath}
holds for all $\phi,\psi\in\Cx^d$, if and only if $M=d\mu\idty$.
\end{Lem}

\proof Inserting the sum defining $\Omega$ we get
\begin{eqnarray*}
&&\mskip-100 mu
    \braket{\phi\otimes\Omega}{(\idty\otimes
          M\otimes\idty)\Omega\otimes\psi}
=\\\mskip60 mu
    &=&\frac1d\sum_{k\ell}\braket{\phi\otimes e_k\otimes e_k}{(\idty\otimes
          M\otimes\idty)e_\ell\otimes e_\ell\otimes\psi}\\
    &=&\frac1d\sum_{k\ell}\braket{\phi}{e_\ell}
            \braket{e_k}{Me_\ell}\braket{e_k}{\psi}
     = \frac1d\braket{\phi}{M^T\psi}\;,
\end{eqnarray*}
which is equal to $\mu\braket\phi\psi$ for all $\phi,\psi$ iff
$M^T=d\mu\idty$. \QED

Consider now the term with index $x\in X$ in the {\em teleportation}
equation~(\ref{telepo}), with $F_x$ and $T_x$ defined via
equations (\ref{Phi2F}), (\ref{U2T}), and (\ref{U2Phi}).
Without loss of generality we set $\rho=\ketbra{\phi_1}{\phi_2}$,
$A=\ketbra{\psi_1}{\psi_2}$. Then
\begin{displaymath}
   {\rm term}_x=\braket{\phi_2\otimes\Omega}{\Phi_x\otimes U_x^*\psi_1}
                  \braket{\Phi_x\otimes U_x^*\psi_2}{\phi_1\otimes\Omega}\;.
\end{displaymath}
The first scalar product can be rewritten by substituting $\Phi_x$
from equation~(\ref{U2Phi}), using equation~(\ref{A2Psi}):
\begin{eqnarray*}
   \braket{\phi_2\otimes\Omega}{\Phi_x\otimes U_x^*\psi_1}
      &=&\braket{\phi_2\otimes((\idty\otimes U_x)\Omega)}
                {((U_x\otimes\idty)\Omega)\otimes\psi_1}
\\    &=&\braket{\phi_2\otimes((U_x^T\otimes\idty)\Omega)}
                {((\idty\otimes U_x^T)\Omega)\otimes\psi_1}
\\    &=&\braket{(\idty\otimes U_x^T\otimes\idty)\phi_2\otimes\Omega}
                {(\idty\otimes U_x^T\otimes\idty)\Omega\otimes\psi_1}
\\    &=&\braket{\phi_2\otimes\Omega}{\Omega\otimes\psi_1}
       = \frac1d\braket{\phi_2}{\psi_1}\;,
\end{eqnarray*}
where at the last equation we used Lemma~\ref{Lem:scprod} with
$\mu=1/d$. Together with a similar computation for the second
scalar product, we get ${\rm
term}_x=d^{-2}\braket{\phi_2}{\psi_1}\braket{\psi_2}{\phi_1}
   =d^{-2}\tr(\rho A)$,
 and equation~(\ref{telepo}) follows by summing over $d^2$ equal
terms.

Similarly, for the {\em dense coding} equation~(\ref{densecod})
we get
\begin{displaymath}
\tr\bigl(\omega (T_x\otimes\id)(F_y)\bigr)
  =\braket{\Omega}{(U_x^*\otimes\idty)\Phi_y}
  \braket{\Phi_y}{(U_x\otimes\idty)\Omega}\;,
\end{displaymath}
i.e., the absolute square of the scalar product
\begin{displaymath}
  \braket{\Omega}{(U_x^*\otimes\idty)\Phi_y}
  =\braket{\Omega}{(U_x^*U_y\otimes\idty)\Omega}
  =\frac1d \tr(U_x^*U_y)
  =\delta_{xy}\;,
\end{displaymath}
where we have used, in turn equation~(\ref{U2Phi}), the maximal
entangledness of $\Omega$ (see equation~(\ref{maxent})), and the
orthogonality of the $U_x$. This completes the proof of the dense
coding property.

\subsection*{Proof of the Implications ``2 $\Longrightarrow$ Rest''}
Let us now assume that a dense coding scheme is given. We have to
conclude that it is of the special form given in
equations~(\ref{O2omega}...\ref{U2Phi}).

Note first that if
 $\omega=\sum_\alpha \lambda_\alpha \omega_\alpha$ ($\lambda_\alpha>0$)
is a mixture of states satisfying the teleportation equation, then
every $\omega_\alpha$ also satisfies it. Hence the assumption is
also satisfied for each pure component $\omega_\alpha$, and we can
first analyze the problem assuming $\omega$ to be pure. In order
to show that $\omega$ indeed is pure, we only have to verify that
the given $F,T$ are consistent only with one pure state. So for
the moment we will assume that $\omega=\ketbra\Omega\Omega$ is
pure.

The next step is a simple general observation on the
coding of classical information on quantum channels, which we
isolate in a Lemma.

\begin{Lem}Let $\K$ be a $D$-dimensional Hilbert space, and
$\sigma_x,F_x\in\B(\K)$, for $x\in X$, a set with $D$ elements.
Suppose that each $\sigma_x$ is a density operator, $F$ is an
observable, and $\tr(\sigma_x F_y)=\delta_{xy}$, for
$x,y=1,\ldots,D$. \\ Then there is an orthonormal basis
$\Phi_x\in\H$ such that
\begin{displaymath}
   \sigma_x=F_x=\ketbra{\Phi_x}{\Phi_x}\;.
\end{displaymath}
\end{Lem}

\proof Let $\Phi_x$ be one of the normalized eigenvectors of
$\sigma_x$ with non-zero eigenvalue. Then since $F_x\leq\idty$,
and $\braket{\Phi_x}{F_x\Phi_x}=1$, $\Phi_x$ must also be an
eigenvector of $F_x$ with eigenvalue $1$. Similarly, for any
$y\neq x$ the $F_x\geq0$, and the normalization $\sum_x F_x=\idty$
forces $F_y\Phi_x=0$. Hence the $\Phi_x$ are orthonormal, and
since their number is the dimension of the space, they must be a
basis. Consequently we have jointly diagonalized the $F_x$ and the
$\sigma_x$, with eigenvalues either $0$ or $1$. \QED

We apply this Lemma with $D=d^2$ and $\sigma_x$
the state after application of $T_x$ to the first factor, i.e.,
$\tr(\sigma_x A)=\tr(\omega (T_x\otimes\id)(A))$. This proves
equation~(\ref{Phi2F}), although it remains to be seen that each
$\Phi_x$ is maximally entangled.

Since the $\sigma_x$ form a maximal set of pure states, there
cannot be a non-zero projection $P$ such that, for all $x\in X$,
\begin{eqnarray*}
   0&=& \tr(\sigma_x(\idty\otimes P))
     =  \tr(\omega (T_x\otimes\id)(\idty\otimes P))\\
    &=& \tr(\omega (\idty\otimes P))
     =  \braket\Omega{(\idty\otimes P)\Omega}\;.
\end{eqnarray*}
Hence $\Omega$ must have {\it full Schmidt rank}. We will need the
consequence that the equation $(A\otimes \idty)\Omega=(A'\otimes
\idty)\Omega$ implies $A=A'$.

Let $T_x(A)=\sum_\alpha K_{x,\alpha}^*AK_{x,\alpha}$ be the Kraus
decomposition of $T_x$. Then the teleportation equation is
\begin{displaymath}
   \sum_\alpha  \abs{\braket\Omega{(K_{x,\alpha}^*\otimes\idty)\Phi_y}}^2
   =\delta_{xy}\;.
\end{displaymath}
Therefore, $\braket{(K_{x,\alpha}\otimes\idty)\Omega}{\Phi_y}=0$
for all $y\neq x$, and for every $x$ there must be constants
$c_\alpha$ such that
\begin{equation}\label{pfpf}
   (K_{x,\alpha}\otimes\idty)\Omega=c_\alpha\Phi_x\;.
\end{equation}
Since $\Omega$ has full Schmidt rank, this implies that all
$K_{x,\alpha}$ are proportional to each other, i.e., that $T_x$
can be written with a single Kraus summand. Of course, the
corresponding $K_x\equiv U_x$ must be unitary, and since both
sides are normalized, equation~(\ref{pfpf})
$\Phi_x=(U_x\otimes\idty)\Omega$, possibly after fixing suitable
phase factors (which influence neither $T_x$ nor $F_x$).

The orthonormality of the $\Phi_x$ translates into $\tr(\rho
U_x^*U_y)=\delta_{xy}$, where $\rho$ is the reduced density
operator of $\omega$. But then Corollary~\ref{noUrhobasis} shows
that $\rho$ must be a multiple of the identity, i.e., $\Omega$ and
each $\Phi_x$ is maximally entangled.

Finally, we have to complete the argument for the purity of
$\omega$ by showing that only one pure state is consistent with
the other data $T,F$, encoded in $U_x$. But this is obvious from
the explicit expression $\Omega=(U_x^*\otimes\idty)\Phi_x$.

\subsection*{Proof of the Implications ``1 $\Longrightarrow$ Rest''}

Let us now assume that a teleportation scheme is given. We have to
conclude that it is of the special form given in
equations~(\ref{O2omega}...\ref{U2Phi}).

The crucial input for this proof is the principle that in quantum mechanics
there is no measurement without perturbation. It enters in the
following form, a corollary of the so-called Radon-Nikodym Theorem
for completely positive maps. We state it here as a Lemma.

\begin{Lem}\label{RaNi}
Let $\H$ be a finite dimensional Hilbert space, and let
$T_\alpha:\B(\H)\to\B(\H)$ be completely positive maps such that
$\sum_\alpha T_\alpha=\id$. Then there are positive numbers $t_\alpha$ such that
$T_\alpha=t_\alpha\id$.
\end{Lem}

\proof For readers less familiar with dilation theory of cp-maps
we include a quick proof based on the Kraus decomposition
$T(A)=\sum_\beta K_\beta^*AK_\beta$, which exists for every
completely positive map. Note that by decomposing each $T_\alpha$
in Kraus form, we get a finer decomposition of $\id$, so we may as
well prove the Lemma for the case that each $T_\alpha$ is of the
form $T_\alpha(A)=K_\alpha^*AK_\alpha$. With $A=\ketbra\psi\psi$,
\begin{displaymath}
   \ketbra{K_\alpha^*\psi}{K_\alpha^*\psi}
   \leq \sum_\alpha\ketbra{K_\alpha^*\psi}{\psi}K_\alpha
   =\ketbra\psi\psi\;.
\end{displaymath}
Hence $K_\alpha^*\psi=\lambda(\psi)\psi$, with a factor
$\lambda(\psi)\in\Cx$. But then {\it every} vector $\psi$ is an
eigenvector of the linear operator $K_\alpha^*$, which is only
possible, if $K_\alpha^*$ is a multiple of the identity.
\QED

A collection of completely positive maps adding up to a normalized
one should be understood as an ``instrument'' in the terminology
of Davies \cite{davies}, i.e., a device which produces classical
measurement results ``$k$'', such that the probability for
obtaining this result {\it and} a response to a subsequent
measurement $F$ on an input state $\rho$ is $\tr(\rho T_k(F)))$.
The channel $\sum_kT_k$ then describes the overall state change,
when the measuring results are ignored. In this language the
hypothesis of the Lemma says that there is no overall state change
through the device, i.e., ``no perturbation'' of the system. The
conclusion is that in that case the output probabilities are
$t_k$, and independent of the input state, i.e., no information
about the system is obtained.

As a first application, we conclude exactly as in the previous
subsection that each convex component of the state $\omega$ again
satisfies the teleportation equation. Hence we can once more
assume that $\omega=\ketbra\Omega\Omega$ is a pure state. The argument that
$\Omega$ is then uniquely determined by the other data, and hence
that $\omega$ is pure is the same as in the dense coding case.

Clearly, this kind of argument is also useful for decompositions
of $T_x$ or $F_x$ into sums of (completely) positive terms. To do
this systematically, fix a maximally entangled unit vector $\Xi$,
so that vectors in $\H\otimes\H$ become expressed as
$\Phi=(A\otimes\idty)\Xi$ for a uniquely determined operator $A$
(see equation~(\ref{A2Psi})). In particular, we can write
$\Omega=(W\otimes\idty)\Xi$, and the Kraus decomposition and
spectral decomposition of each $F_x$ in the form
\begin{eqnarray}
  T_x(A)&=&\sum_\alpha K_{x,\alpha}^*AK_{x,\alpha}\\
  F_x   &=&\sum_\beta (A_{x,\beta}\otimes\idty)\ketbra\Xi\Xi
                       (A_{x,\beta}\otimes\idty)^*\;.
\end{eqnarray}
Inserting this into the teleportation equation~(\ref{telepo})
we find a sum over
$x,\alpha,\beta$, in which each term represents a completely
positive operator, and which sum up to the identity. Hence by
Lemma~\ref{RaNi}, each
term has to be multiple of the identity,
$\widetilde\mu_{x,\alpha,\beta}\id$, say. This can be
written in terms of scalar products, if we take
$\rho=\ketbra{\phi_1}{\phi_2}$ and $A=\ketbra{\psi_1}{\psi_2}$:
\begin{eqnarray*}\label{scsc}
  \widetilde\mu_{x,\alpha,\beta}\ \tr(\rho\ A)
  &=& \widetilde\mu_{x,\alpha,\beta}\
       \langle\phi_2,\psi_1\rangle
       \langle\psi_2,\phi_1\rangle\\
  &=&\bigl\langle{\phi_2\otimes\Omega}, (A_{x,\beta}\otimes\idty
         \otimes K_{x,\alpha}^*)\Xi\otimes\psi_1\bigr\rangle\\
   &&\qquad\bigl\langle\Xi\otimes\psi_2,\
             (A^*_{x,\beta}\otimes\idty\otimes
      K_{x,\alpha}){\phi_1\otimes\Omega}\bigr\rangle\;.
\end{eqnarray*}
Note that the two scalar products on the right hand side are complex conjugates of each
other apart from a swapping of the arguments $(\phi_2,\psi_1)$ and
$(\psi_2,\phi_1)$, which exactly matches the variable pairing on
the left hand side. Since the equation is to hold for arbitrary
vectors $\phi_1,\phi_2,\psi_1,\psi_2$, we can hold one pair fixed
and find that
\begin{equation}\label{halftelepo}
  \bigl\langle{\phi_2\otimes\Omega}, (A_{x,\beta}\otimes\idty
         \otimes K_{x,\alpha}^*)\Xi\otimes\psi_1\bigr\rangle
  ={\mu_{x,\alpha,\beta}}\
       \langle\phi_2,\psi_1\rangle\;,
\end{equation}
where $\mu_{x,\alpha,\beta}$ is a factor determined in terms of
$\widetilde\mu$, and the scalar products involving
$(\psi_2,\phi_1)$.  With
$\Omega=(W\otimes\idty)\Xi$, and equation~(\ref{A2Psi}) we get
\begin{eqnarray*}\label{halftelepo1}
  &&\mskip-100 mu \bigl\langle{\phi_2\otimes\Omega}, (A_{x,\beta}\otimes\idty
         \otimes K_{x,\alpha}^*)\Xi\otimes\psi_1\bigr\rangle\\
\ \mskip60 mu\
&=&\ \langle{\phi_2\otimes(\idty\otimes K_{x,\alpha})\Xi},
           (\idty\otimes W^*\otimes\idty)
           \bigl((A_{x,\beta}\otimes\idty)
        \Xi\otimes\psi_1\bigr)\bigr\rangle\\
&=&\ \langle{\phi_2\otimes(K_{x,\alpha}^T\otimes\idty)\Xi},
           (\idty\otimes W^*\otimes\idty)
           \bigl((\idty\otimes A^T_{x,\beta})
        \Xi\otimes\psi_1\bigr)\bigr\rangle\\
&=&\ \langle{\phi_2\otimes\Xi},
           (\idty\otimes \overline{K_{x,\alpha}}\; W^*A^T_{x,\beta}\otimes\idty)
           \Xi\otimes\psi_1\bigr\rangle\\
&\equiv&{\mu_{x,\alpha,\beta}}\
           \langle\phi_2,\psi_1\rangle\;,
\end{eqnarray*}
where we have used the notation $\overline{K}=(K^*)^T$ for the
matrix element-wise complex conjugation in the Schmidt basis
belonging to the maximally entangled state $\Xi$. Since the above
equation holds for all $\phi_2$ and  $\psi_1$,
Lemma~\ref{Lem:scprod} implies that
\begin{equation}\label{teleOperator}
   \overline{K_{x,\alpha}}\; W^*A^T_{x,\beta}
      =d\;\mu_{x,\alpha,\beta}\ \idty\;,
\end{equation}
for all $x,\alpha,\beta$.

Let us say that a label $x\in X$ {\it contributes} to
teleportation, if the corresponding term in the teleportation
equation does not vanish for all $\rho$ and $A$. This is
equivalent to saying that  for some $\alpha,\beta$
the factor $\mu_{x,\alpha,\beta}$ is non-zero.
For such triples $(x,\alpha,\beta)$ all
three operators on the left hand side of equation~(\ref{teleOperator})
have to be invertible.

Now since there has to be at least one contributing label, $W$ has
to be non-singular, which means that $\Omega$ has {\it full
Schmidt rank}. Equivalently, the reduced density operator
$\omega_1$ for the first factor has no zero eigenvalues. From this
we conclude that the non-contributing labels are precisely those
for which $F_x=0$. Indeed, we may set $A=\rho=\idty$, and use the
normalization of $T_x$ to find
\begin{displaymath}
 0= \tr\bigl((\idty\otimes\omega)( F_x\otimes\idty)\bigr)
  = \tr\bigl((\idty\otimes\omega_1) F_x\bigr)\;
\end{displaymath}
Since $F_x\geq0$, and $\idty\otimes\omega_1$ has only strictly
positive eigenvalues, this implies $F_x=0$.

Now let $x$ be a contributing index, and choose some triple $(x,\alpha,\beta)$
with $\mu_{x,\alpha,\beta}\neq0$.
If we now look at equation~(\ref{teleOperator}) for
triples $(x,\alpha',\beta)$ with arbitrary $\alpha'$, we get
$\overline{K_{x,\alpha'}}
   =(\mu_{x,\alpha',\beta}/\mu_{x,\alpha,\beta})\overline{K_{x,\alpha}}$,
i.e., all Kraus operators of $T_x$ are proportional, and hence
$T_x$ can be written with a single Kraus summand,
$T_x(A)=U_x^*AU_x$, with a unitary $U_x$.

Similarly, we find that all $A_{x,\beta'}$ are
proportional, which means that $F_x=\ketbra{\Phi_x}{\Phi_x}$ with
$\Phi_x=(A_x\otimes\idty)\Xi$.

We can now apply Lemma~\ref{l.complete} to these vectors $\Phi_x$,
setting $\Phi_x=0$ for non-contributing labels. The conclusion is that
the $\Phi_x$ are an orthonormal basis. In particular, all indices
do contribute after all.

Equation~(\ref{teleOperator}) and the unitarity of $U_x$ allow us
to express $A_x$ in terms of $U_x$:
\begin{equation}\label{K2A}
  A_x=d\mu_x\ U_x\overline{W}^{-1}\;
\end{equation}
Orthonormality of the $\Phi_x$ becomes
\begin{equation}\label{WW-1}
  \delta_{xy}= \frac1d\tr(A_x^*A_y)
     =d \overline{\mu_x}\mu_y
       \tr(U_y \; \overline{W}^{-1}(\overline{W}^{-1})^*\;U_x^*)\;.
\end{equation}
For $x=y$ we find that $\abs{\mu_x}^2$ is independent of $x$,
hence the operators $(\overline{\mu_x}/\abs{\mu_x})U_x^*$ are
unitary, and satisfy the hypothesis of Corollary~\ref{noUrhobasis}
with $\rho$ a positive multiple of
$\overline{W}^{-1}(\overline{W}^{-1})^*$. Hence this operator is a
multiple of the identity, $W$ is unitary up to a factor, and
$\Omega=(W\otimes\idty)\Xi$ is maximally entangled. Moreover,
we see from equation~(\ref{WW-1}) and the $U_x$ form a unitary basis.

Since $\Xi$ was an arbitrary maximally entangled vector, we may
just as well take $\Xi=\Omega$, so equation~(\ref{K2A}) holds with
$W=\idty$. Hence, $\Phi_x=c\;(U_x\otimes\idty)\Omega$, where $c$
is a factor which has to be of modulus $1$, because $\Omega$ and
$\Phi_x$ are normalized, and $U_x$ is unitary, and which can be
chosen to be $1$ by adjusting the phase of $\Phi_x$. This
completes the proof.

\section{Constructing bases of unitaries}

It is not a priori clear that bases of unitary operators should
exist in any dimension. Indeed, the system
equation~(\ref{unibasis}) of equations is formally overdetermined,
according to the following rough dimension count. The variables in
this system are the unitaries $U_x$, each of which we can take in
the $(d^2-1)$-dimensional manifold $SU_d$, i.e., with
$\det{U_x}=1$, by fixing a phase factor. Since the transformations
$U_x\mapsto V_1U_xV_2$, for arbitrary $V_1,V_2\in SU_d$ leave the
set of solutions invariant, we may fix $U_1=\idty$, and take $U_2$
diagonal without loss of generality. This reduces the number of
variables to $(d-1)+(d^2-2)(d^2-1)$. On the other hand,
orthogonality introduces one complex constraint for every pair
$x\neq y$. None of these is trivially satisfied due to the special
choices we made, so we have to take $d^2(d^2-1)$ constraints into
account. This leaves, formally,
\begin{displaymath}
\text{\#variables-\#equations}
       =-(d-1)(2d+1)<0\;.
\end{displaymath}
Of course, we know that this count is somehow too crude, because,
after all, many inequivalent unitary bases are constructed below.
But it is not so easy to spot the dependences among the
constraints. Note also that the dimension count is essentially the
same for bases orthogonal with respect to a weight $\rho\neq
d^{-1}\idty$, but in that case Corollary~\ref{noUrhobasis} shows
that there is no solution at all.

In order to describe the best known construction for unitary bases
\cite{tqu}, let us introduce some terminology. We say that a
(single) unitary matrix is of {\em shift and multiply} type, if it
is the product of a permutation operator and a diagonal unitary.
In other words, every row or column contains $(d-1)$ zero entries,
and one entry of modulus $1$. The bases we will construct not only
have the property that each element is of this type, but also that
the $d^2$ values for $x$ can be split into $d$ options for
``shift'' and $d$ options for ``multiply''.

\begin{Def}\label{s+m} A {\bf shift and multiply basis} of unitary
matrices in $\Cx^d$ is a collection of $d^2$ unitary operators
$U_{ij}$, $i,j\in I_d\equiv\{1,\ldots,d\}$, satisfying the
orthogonality relation
$\tr(U_{ij}^*U_{k\ell})=d\;\delta_{ik}\delta_{j\ell}$, and
 acting on the basis vectors $\ket k$ as
\begin{equation}\label{e:s+m}
  U_{ij}\ \ket k=H^j_{ik}\ \ket{\lambda(j,k)}\;,
\end{equation}
where the $H^j_{ik}$ are complex numbers, and
$\lambda:I_d\times I_d\to I_d$.
\end{Def}

\begin{Pro}
The parameters  and $\lambda:I_d\times I_d\to I_d$
define a shift and multiply basis of unitary matrices if and only
if the following two conditions are satisfied
\begin{enumerate}
\item Each $H^j$ is a {\bf Hadamard matrix}, i.e.
$\abs{H^j_{ik}}=1$ for all $i,k$, and
$H^j(H^j)^*=d\;\idty$.
\item $\lambda$ is a {\bf Latin square}, i.e., the maps
 $k\mapsto\lambda(k,\ell)$ and  $k\mapsto\lambda(\ell,k)$ are
injective for every $\ell$.
\end{enumerate}
\end{Pro}

\proof For $U_{ij}$ to be unitary, it is necessary and sufficient
that the $H^j_{ik}$ are phases, and that $k\mapsto\lambda(j,k)$ is
injective (hence bijective) for every $j$. For the orthogonality
we have to evaluate
\begin{displaymath}
  \tr(U_{ij}^*U_{i'j'})=\sum_k \overline{H^j_{ik}}H^{j'}_{i'k}
      \braket{\lambda(j,k)}{\lambda(j',k)}\;.
\end{displaymath}
We consider first the case $j=j'$. Then the scalar products in the
sum are all equal to $1$, and equating this expression to
$\delta_{ii'}$ we find that $H^j$ is Hadamard.

Now let $j\neq j'$, and consider the ``coincidence set''
 $C=\{k\mid\lambda(j,k)=\lambda(j',k)\}$.
Then orthogonality requires, for every $i,i'$, that
\begin{equation}
  0 = \sum_{k\in C }\overline{H^j_{ik}}H^{j'}_{i'k}
    = \sum_{k=1}^d  H^{j'}_{i'k}\ \chi_C(k)\ (H^{j*})_{ki}
    = (H^{j'}\chi_C H^{j*})_{i'i}\;,
\end{equation}
where $\chi_C(k)=1$ for $k\in C$, and zero otherwise, and in the
last line $\chi_C$ denotes the projection $\chi_C\ket
k=\chi_C(k)\ket k$. But since $H^{j'}$ and $H^j$ are Hadamard, and
in particular invertible, this implies $\chi_C=0$. Hence $C$ is
empty, and the second injectivity of $\lambda$ is proved. \QED

In order to construct unitary bases of this form, we must now
construct Hadamard matrices and Latin squares of the appropriate
dimension. For both of these tasks there is a rich literature, and
below we will give a brief summary on what is known for each.

It is useful to note that each of the structures `unitary bases',
`Hadamard matrices', and `Latin squares' has a natural notion of {\it
equivalence}, and to some extent these equivalences are related.
We call two unitary bases $U,U'$ equivalent, if
$U'_x=V_1U_{x'}V_2$, for some unitaries $V_1,V_2$, and a
re-labelling $x\mapsto x'$. Hadamard matrices are called
equivalent, if one is obtained from the other by permuting
rows or columns, or multiplying rows or columns with phases.
Finally, a Latin square $\lambda:I_d\times I_d\to I_d$ is
equivalent to any other obtained by applying a permutation on each
of the three copies of $I_d$ involved. In each case there are also
discrete transformations, such as transposition or complex
conjugation (where applicable). It should be noted that replacing
each $H^j$ by an equivalent one, typically only leads to an
equivalent unitary basis, if the equivalence operation is the same
for each $j$. With $j$-dependent equivalence transformations it is
possible to construct inequivalent unitary bases in $d=3$,
although in this dimension there is only one Hadamard matrix and
only one Latin square -- up to equivalence. Of course, in
$d=2$ all three structures, including  the unitary bases
are unique up to equivalence \cite{tqu}.  The unique unitary basis
is then given by the three Pauli matrices and the identity and,
of course generates via the Theorem~1, the usual two qubit examples of
teleportation and dense coding.

For each of the three structures we furthermore have an obvious
notion of {\it tensor product}, allowing the construction of a
unitary basis (resp. a Hadamard matrix, or Latin square) in
dimension $d=d_1d_2$, if counterparts in dimension $d_1$ and
dimension $d_2$ are given.

In order to show that unitary bases exist in any dimension it is
easiest to use group theory based constructions: the Latin square
can be taken as the multiplication table of any group of order
$d$, for example the cyclic group. The Hadamard matrix can be
taken as the matrix implementing the Fourier transform on an
abelian group of order $d$, the standard example being given once
again by the cyclic group of order $d$. Thus
$H_{k\ell}=\exp(\frac{2\pi i}dk\ell)$, where $k$ and $\ell$ are
taken modulo $d$. If we combine these data into a unitary basis we
get an instance of what we propose to call a {\it unitary basis of
group type} (``nice error basis'' in \cite{Knill}). These are
orthonormal unitary bases with the additional property that the
operator product of any two elements is a third, up to a phase.
That is to say, the index set $X$ is a group, and
\begin{equation}\label{Ugroup}
  U_xU_y=\mu(x,y)U_{xy}\;,
\end{equation}
with $\abs{\mu(x,y)}=1$. In the special case of an abelian group
$X$ this is a discrete version of {\it Weyl systems} of unitary
operators, named after their continuous variable counterpart,
well-known from quantum optics and non-relativistic ``phase
space'' quantum mechanics.

{\em Latin squares} are not completely classified, nor does there
seem to be a realistic hope to do so. A standard work on the
subject is \cite{denkeed}, a useful net resource is \cite{ritter}.
Counts of squares are usually done for ``normalized squares'', in
which the first row and column are in natural order, thus
eliminating some trivial freedom. In $d=5$ Euler counted $56$ of
these, but only $2$ are inequivalent, because the symbols
themselves can also be permuted. Counts of normalized squares have
now gone all the way up to $d=10$, but are no longer done by hand
(there are roughly $7.5\times10^{24}$ \cite{McKay}). It is also
clear from these numbers that group based constructions exhaust
only a tiny fraction of the possible unitary bases.

{\em Hadamard matrices} are also a standard subject in coding
theory. However, usually only the real case (orthogonal matrices
with entries $\pm1$) is considered. It is easy to see that real
Hadamard matrices exist only in dimension two and multiples of
four. Again, the possibilities for such designs by far exceed the
group based possibilities (the characters of an abelian group are
real only if $d=2^n$). A standard reference is \cite{agaian}.

For complex Hadamard matrices the Fourier matrices show that there
is no constraint on dimension. The uniqueness in $d=3$ is easy to
get. The general form in $d=4$ is, up to equivalence
\begin{equation}\label{Hada4}
  \left(\begin{array}{cccc}
  1& 1&  1& 1\\
  1& 1& -1&-1\\
  1&-1&  u&-u\\
  1&-1& -u& u
  \end{array}\right)\ ,
\end{equation}
where $u$ is an arbitrary phase. For
$u=1$ this is equivalent to the Fourier matrix of ``Klein's Four
Group'', the product of two copies of the two-element group,
and for $u=i$ it is equivalent to the Fourier matrix of the cyclic group.
The possibility of embedding the cyclic group Fourier matrix into
a higher dimensional manifold can be generalized to arbitrary
composite numbers $d=pq$: whenever $V_{k\ell}$ is a matrix of
phases satisfying the periodicity conditions
$V_{k,\ell}=V_{k+p,\ell}=V_{k,\ell+q}$, we get a Hadamard matrix
as
\begin{equation}\label{Hadapq}
  H_{k\ell}=V_{k\ell} \exp\left(\frac{2\pi i}d\;k\ell \right)\;.
\end{equation}

One might conjecture from this that for prime orders $d$ the
Hada\-mard matrix is unique. This problem was discussed by
Haagerup \cite{haag} on the basis of a completely different
motivation (theory of von Neumann algebras). There it is shown
that $d=5$ there is uniqueness, but for $d=7$ there are at least
$5$ solutions. For some primes, uncountably many inequivalent
Hadamard matrices are known.

\end{document}